# Signal Fluctuation Sensitivity:
# an improved metric for optimizing detection
# of resting-state fMRI networks

D.J. DeDora[1*], S. Nedic[1*], P. Katti[1], S. Arnab[1], L.L. Wald[2,3,4], A. Takahashi[5], K.R.A. Van Dijk[2,6], H.H. Strey[1], L.R. Mujica-Parodi[1,2,3,5]

*Authors contributed equally to the submitted work.

1. Department of Biomedical Engineering, Stony Brook University School of Medicine, Stony Brook, NY 11794 USA

2. Department of Radiology, A. A. Martinos Center for Biomedical Imaging, Massachusetts General Hospital, Charlestown, MA 02129 USA

3. Harvard Medical School, Boston, MA 02115 USA

4. Harvard-MIT Division of Health Sciences and Technology, Massachusetts Institute of Technology, Cambridge, MA 02139 USA

5. McGovern Institute for Brain Research at MIT, Massachusetts Institute of Technology, Boston, MA 02139

6. Harvard University Department of Psychology, Center for Brain Science, Cambridge, MA 02138 USA

Please address all correspondence to:

Dr. LR Mujica-Parodi
Director, Laboratory for Computational Neurodiagnostics
Associate Professor, Department of Biomedical Engineering
Bioengineering Building, Room 119
Stony Brook University
Stony Brook, NY 11794-5281
Telephone: 917-669-3934
Email: Lilianne.Strey@stonybrook.edu

Abstract

Task-free connectivity analyses have emerged as a powerful tool in functional neuroimaging. Because the cross-correlations that underlie connectivity measures are sensitive to distortion of time-series, here we used a novel dynamic phantom to provide a ground truth for dynamic fidelity between *blood oxygen level dependent* (BOLD)-like inputs and fMRI outputs. We found that the *de facto* quality-metric for task-free fMRI, *temporal signal to noise ratio* (tSNR), correlated inversely with dynamic fidelity; thus, studies optimized for tSNR actually produced time-series that showed the greatest distortion of signal dynamics. Instead, the phantom showed that dynamic fidelity is reasonably approximated by a measure that, unlike tSNR, dissociates signal dynamics from scanner artifact. We then tested this measure, *signal fluctuation sensitivity* (SFS), against human resting-state data. As predicted by the phantom, SFS—and not tSNR—is associated with enhanced sensitivity to both local and long-range connectivity within the brain's default mode network.

Introduction

Unprecedented investment in functional neuroimaging has ushered in a new era of brain research, in which fMRI's original role in mapping the areas of the brain most 'active' under a task, now includes task-free characterization of brain connections and circuits. This evolution implies a fundamental—and yet largely unacknowledged—shift in how we understand *signal* versus *noise*.

During fMRI's first decade, researchers almost exclusively used stimulus presentation to evoke *blood oxygen level dependent* (*BOLD*) activity in subjects. To identify the relationship between different brain regions and their functional roles, tasks included one or more experimental conditions (*tasks*), as well as a baseline measure absent of stimuli (*rest*). FMRI time-series were then fitted to the expected hemodynamic shape for each condition (*canonical hemodynamic response function*, or *HRF*). Once fitted, trials for each condition were averaged and used to statistically compare hemodynamic amplitudes for each condition (*contrasts*) across subjects. Contrasts that met statistical thresholds were then represented as activity, producing activation maps. Importantly, the fitting, averaging, and subtraction approach used to analyze task-based data was designed to distinguish between time-series fluctuations originating from two sources. On the one hand, it amplified *fluctuations of interest* corresponding to task-based activation and that therefore were correlated with the experiment's design matrix. On the other hand, it suppressed *fluctuations of nuisance* that corresponding to (scanner, physiological) artifact and that therefore were independent of the experiment's design matrix.

In the late 1990's several influential papers[1-4] showed for the first time that the brain showed strong and reliable correlations between fMRI time-series even in the absence of a well-defined task (*resting-state connectivity*); more recently, the relationship between correlation-derived networks, and the neuronal events that underlie them, has been identified using fMRI acquired simultaneously with electrophysiological recordings of local field potentials[5]. The fMRI community quickly responded, and today task-free *connectivity* studies—which map connections between brain nodes as defined by correlations between their time-series—comprise over 20% of human neuroimaging studies published every year[6]. Connectivity analyses include not only those obtained by correlations with a pre-defined region (*seed-based*) but also those that describe graph-theoretic features of the functional connectome (*complex network analyses*)[7]. Together, connectivity studies have contributed a wealth of new human brain data on aging[8], psychiatric[9] and neurological[10] disorders, and injury[11]. Resting-state fMRI protocols are easily standardized, require minimal patient compliance, and permit exploratory analyses; as such, they would appear to be well positioned for both clinical neurodiagnostics as well as large-scale international bio-repositories established for epidemiological research.

However, the transition from *activation maps* to *connections between nodes* not only produced a conceptual shift with respect to the role of functional neuroimaging, but also increased dependence upon time-series power spectra (see Online Methods). The standard measure for establishing the quality of task-based data has been the *contrast-to-noise ratio (CNR)*, defined as the contrast (mean activation level acquired during task minus the mean activation level acquired during rest) divided by the standard deviation of the time-series[12]. For task-free designs however, CNR cannot be computed, and thus normally is replaced by the *temporal signal-to-noise ratio* (*tSNR*), defined as the mean of the time-series divided by its standard deviation[13]. Intuitively, both CNR and tSNR compare the amplitude of a signal against a background of undesired physiological, thermal, and scanner noise present in all fMRI studies. This manner of conceptualizing what is 'signal' versus what is 'noise' makes perfect sense within the context of activation maps, in which a task activates the brain reliably more under one condition (signal) than another (noise)[14]. However, for task-free analyses, the 'baseline' fluctuations themselves also include the 'signal.' Thus, for most task-free analyses, tSNR would appear to do exactly the opposite of what one would wish, as it penalizes sensitivity to the fluctuations (i.e., the standard deviation of the time-series) upon which experimental results are also based. Indeed, several recent studies have reported little correspondence between resting-state tSNR and the detection of stable functional networks[15-18].

For task-free analyses, rather than relegate time-series fluctuations to the category of noise as per tSNR, we want to—as with task-based analyses—functionally distinguish between fluctuations of interest that are *neurobiologically significant* (e.g., emanating from BOLD signal consequent to neuronal response) from fluctuations of nuisance that are *neurobiologically insignificant* (e.g., physiological, scanner, and motion artifact). The dissociation between the two can be characterized by *signal fluctuation sensitivity (SFS)*, which we define at a single-voxel level as:

**Eqn. 1**

$$SFS_{voxel} = \frac{\mu_{ROI}}{<\mu_{global}>} \times \frac{\sigma_{ROI}}{<\sigma_{nuisance}>}$$

In the first term, the numerator consists of the mean signal ($\mu$) of a time-series acquired from a voxel in the *region of interest* (*ROI*). For the denominator, we average over all voxel-specific signal for the entire brain (*global*). The first term ensures that SFS decreases for regions with signal drop out, while remaining unit-less (as with tSNR). In the second factor, the numerator consists of the standard deviation ($\sigma$) of the time-series acquired from the voxel of interest. For the denominator, we average over all voxel-specific $\sigma$ from a region in which BOLD signals are not expected, but in which physiological, scanner, and motion artifacts are still present (*nuisance*). Prior work suggests that time-series obtained from cerebrospinal fluid (CSF) meet criteria for the *nuisance* denominator[19]. SFS for a region of interest is then computed by averaging voxel-specific SFS values over all voxels in the region ($SFS_{ROI} = <SFS_{voxel}>_{ROI}$). In order to more easily compare SFS with tSNR, we scale them comparably by multiplying SFS values by 100.

We define *dynamic fidelity* as the degree to which fMRI accurately captures true BOLD fluctuations. In order to test our hypothesis that SFS should reflect dynamic fidelity of time-series more accurately that tSNR, we first needed to know the 'ground truth' for those fluctuations. To access that ground truth, we designed and constructed a *dynamic phantom*, which provides user-controlled—and thus known—dynamic BOLD-like inputs to which fMRI-derived outputs can be compared. We then tested the impact of dynamic fidelity, as defined by our phantom, in predicting detection sensitivity to functional connectivity in human data across three different sets of acquisition parameters, chosen to represent a breadth of realistic optimization strategies utilized within the neuroimaging field for human connectivity studies.

Results

*Dynamic Phantom Design*
Our dynamic phantom exploits the fact that the magnetic susceptibility of agarose gel is concentration-dependent; thus, varying the concentration of agarose gel present within a voxel over time produces changes in T2* that we experimentally tuned to amplitudes (~1%) typically observed with BOLD[20]. The dynamic phantom is constructed from two concentric cylinders coupled with a pneumatic motor and fiber optic feedback system (**Fig. 1a**). The outer cylinder contains a 'baseline' agarose gel (2.27% w/w), while the inner cylinder is longitudinally divided with both (*i*) a baseline gel matching the outer cylinder and (*ii*) an 'active' gel with slightly lower concentration of agarose (2.21% w/w), which produces slightly greater fMRI signal than the baseline gel. The longitudinally divided inner cylinder rotates about its long axis via a novel fMRI-compatible pneumatic motor to drive rotation of the inner cylinder (**Fig. 1a-b**). By averaging time-series across a *region of interest* (ROI) that, over time, includes different proportions of the two concentrations, we were able to reproduce the effect of a concentration gradient in producing smooth fMRI time-series (**Fig. 1c**). The dynamic phantom receives image acquisition signals from the fMRI, and rotates only between image acquisitions to avoid motion artifacts. As the dynamic phantom rotates, position is monitored continuously through a fiber-optic feedback system. As the fMRI acquires each image, the dynamic phantom reads out its position, which serves as a 1:1 'input' for input-output mapping. The dynamic phantom can be programmed to produce fMRI signals that follow any dynamic input, including those expected for task-generated event-related and block designs, as well as resting-state (e.g., pink-noise) fluctuations. As shown in **Fig. 1d**, the dynamic phantom produces tightly controlled changes in fMRI signal without motion artifacts, and therefore can provide a ground-truth upon which to establish the degree to which SFS and tSNR reflect fMRI's dynamic fidelity to the original BOLD signal.

*Dynamic Phantom Assessment of SFS vs tSNR in Predicting Dynamic Integrity*
We programmed the dynamic phantom to mimic resting-state oscillations observed in human fMRI[21] (**Fig. 2a**), and scanned the dynamic phantom under three different sets of acquisition parameters. *Acquisition A* represents what would normally be considered to be the standard for typical resting-state studies, using a 3T scanner with 32-channel head coil and 2000ms temporal-resolution (TR). *Acquisition B* uses a set of parameters that were specifically designed for resting-state connectivity analyses as part of the *Human Connectome Project*. These include a 3T scanner that increases the temporal-resolution to 1080ms in order to achieve greater sensitivity to fluctuation dynamics; to compensate for signal loss associated with accelerated scanning, *Acquisition B* uses a custom-built 64-channel head coil. *Acquisition C* pushes even further than *Acquisition B* in optimizing over temporal resolution (802ms). *Acquisition C* retains the 32-channel head coil, but compensates for signal loss associated with accelerated scanning by increasing the field strength to 7T. In each scanner, we scanned the dynamic phantom for 10 minutes under each acquisition paradigm optimized for human studies, as well as at two other TRs comparable to those previously optimized for the other two scanners. Thus, in total we performed nine scans (three scanners × three TRs each) with the dynamic phantom; scanners and scan parameters used for each session are provided in **Table 1**. For human data, we also acquired T1-weighted structural images and B0 field maps for correction of EPI data (see Online Methods).

We then computed dynamic fidelity, SFS, and tSNR on raw data acquired from the dynamic phantom. Standard deviations were computed after voxel-wise removal of linear and quadratic trends. The dynamic phantom is longitudinally divided into four chambers, and rotates about the long axis orthogonally to the main field. For our *region of interest*, we extracted the average time-series from each the four quadrants of the inner cylinder with an automated masking procedure, and repeated this for six slices positioned in the center of the dynamic phantom (n = 24 time-series per scan). *Dynamic fidelity* was defined as the correlation between user-defined *dynamic inputs*, provided by the phantom rotation, and *dynamic outputs* acquired from the scanner in the region of interest. To compute the *nuisance* term within SFS (analogous to CSF in humans), we extracted fluctuations acquired from the outer cylinder in these six slices, which includes only inactive voxels.

Dynamic fidelity *directly* correlated with SF*S* for each of the nine scans (**Fig. 2b**; **Table 2**; median $r$ = 0.67) and *inversely* correlated with tSNR for each of the nine scans (**Fig. 2b**; **Table 2**; median $r$ = –0.63). Thus, when the scanner was most sensitive in capturing dynamic inputs, SFS was maximized while tSNR was minimized, and vice-versa. Researchers typically optimize acquisition parameters for fMRI connectivity studies by trying to maximize tSNR. Yet doing so would appear to produce the greatest amount of distortion for the BOLD fluctuations upon which connectivity results are based. Thus, we thus tested the implications of our dynamic phantom results for human connectivity studies.

## *Human Subjects Assessment of SFS vs. tSNR in Detecting Resting-State Connectivity*

We calculated SFS and tSNR in human neuroimaging data acquired using *Acquisitions A*, *B*, and *C* (restricting our analyses to the TR originally, and independently, optimized for each scanner), and assessed the utility of each in predicting detection sensitivity to resting-state network features. Human data were preprocessed according to standard methods, including the SPM8 preprocessing pipeline (see Online Methods); to gauge the impact of spatial smoothing, we calculated all values both with and without this step. After preprocessing, we used MATLAB to compute SFS and tSNR as per **Eqn. 1**. For resting-state connectivity, we computed three commonly used measures. The first was the between-voxel measure of local connectivity, *regional homogeneity* (*ReHo*)[22]. The second was the within-voxel *amplitude of low-frequency fluctuations* (*ALFF*)[23], which is thought to underlie resting-state connectivity[24]. The third was *long-range connectivity* between two nodes of the default model network[25]: the *medial prefrontal cortex* and the *posterior cingulate cortex* (mPFC-PCC). DMN regions were defined as 10mm radius spheres centered upon previously established coordinates[26]. Long-range connectivity forms the basis for graph theoretic/complex network analyses[7] used within the fMRI field.

To test the degree to which SFS and tSNR were sensitive to well-established resting-state features, we computed correlations between SFS and ReHo, ALFF, and long-range connectivity; as well as tSNR and ReHo, ALFF, and long-range connectivity. ReHo and ALFF were computed for voxels within the well-established default mode network, comprised of the *medial prefrontal cortex*, *posterior cingulate cortex*, and *bilateral parietal cortices*. Long-range connectivity focused upon the two-node MPFC-PCC connection, which we found to be reliable across subjects within our dataset (33 out of 36 subjects showed significant positive correlation between mean time series from these two regions). For networks that include two or more nodes, we used the minimum SFS or tSNR for each mPFC-PCC pair (see Online Methods).

We first tested whether SFS and tSNR would predict local connectivity (ReHo) at a single-subject level. Without smoothing, region-specific correlations within the *default mode network* showed robust positive relationships between SFS and ReHo (median $r = 0.53$: 96% $p < 0.05$, 95% $p < 0.01$, 94% $p < 0.001$; by acquisition set: $r_A = 0.54$, $r_B = 0.51$, $r_C = 0.54$; see **Fig. 3b** for median across subjects and default mode network regions). In contrast, the correlation between tSNR and ReHo was either non-significant or significant but negative within most subjects' default mode network regions (median $r = -0.24$: 80% $p < 0.05$, 76% $p < 0.01$, 68% $p < 0.001$—even using the most liberal threshold of $p < 0.05$, only 11% of all correlations were positive between tSNR and ReHo; by acquisition set: $r_A = -0.42$, $r_B = -0.20$, $r_C = -0.06$; see **Fig. 3c** for median across subjects and default mode network regions). Smoothing only magnified this effect. After smoothing, SFS positively correlated with ReHo (median $r = 0.68$: 98% $p < 0.05$, 98% $p < 0.01$, 98% $p < 0.001$; by acquisition set: $r_A = 0.69$, $r_B = 0.72$, $r_C = 0.55$) and tSNR negatively correlated with ReHo (median $r = -0.62$: 97% $p < 0.05$, 97% $p < 0.01$, 97% $p < 0.001$; by acquisition set: $r_A = -0.72$, $r_B = -0.60$, $r_C = -0.57$).

While ReHo is a measure of between-voxel local connectivity, ALFF is a single-voxel measure that estimates the total power of the low frequency component of an fMRI signal. Thus, we expected the relationship between ALFF and SFS (both single voxel measures) to be even more robust than the relationship between SFS and ReHo. Indeed, SFS strongly correlated with ALFF (median $r = 0.82$, all $p$'s $< 0.001$; by acquisition set: $r_A = 0.90$, $r_B = 0.71$, $r_C = 0.77$), whereas tSNR was negatively correlated with ALFF (median $r = -0.70$, all $p$'s $< 0.001$; by acquisition set: $r_A = -0.82$, $r_B = -0.65$, $r_C = -0.58$). Again, smoothing magnified this effect for both SFS (median $r = 0.93$, all $p$'s $\ll 0.001$; by acquisition set: $r_A = 0.94$, $r_B = 0.92$, $r_C = 0.93$), and tSNR (median $r = -0.84$, all $p$'s $\ll 0.001$; by acquisition set: $r_A = -0.86$, $r_B = -0.83$, $r_C = -0.84$).

As a measure of long-range connectivity, we tested SFS and tSNR against the default mode network's MPFC-PCC connection (Fisher-z normalized) across our three datasets (N=36). Consistent with the other connectivity measures, SFS positively correlated with MPFC-PCC connectivity ($r_{A,B,C} = 0.61$, $p = 8.65 \times 10^{-5}$) and tSNR negatively correlated with MPFC-PCC connectivity ($r_{A,B,C} = -0.73$, $p = 4.46 \times 10^{-7}$). As with previous measures, smoothing did not qualitatively change the results for either SFS ($r_{A,B,C} = 0.40$, $p = 0.015$) or tSNR ($r_{A,B,C} = -0.70$, $p = 1.67 \times 10^{-6}$).

*SFS and tSNR Values Between Acquisition Sets*
The purpose of sensitivity metrics, for any measurement, is to provide accurate feedback by which parameters can be tuned to optimize performance, as well as to aid in the interpretation and artifact-correction of results. Our three representative acquisition strategies illustrate clearly the practical importance of using SFS rather than tSNR when optimizing fMRI studies for task-free analyses, and therefore dynamic fidelity. We compared SFS and tSNR values between acquisition paradigms for the default mode network, subcortical regions critical to the emotion and reward circuits, and global gray matter. Because human studies normally utilize smoothing, and because our previous analyses (above) showed comparable results for smoothed and unsmoothed results, **Fig. 4** results include only 4-mm smoothed data. To directly compare SFS and tSNR values between acquisition sets, we extracted average values from the four regions of the *default mode network* and three subcortical regions (*amygdala*, *caudate*, *hippocampus)*, as well as *average subcortical* and *average gray matter*. Each subcortical region was defined from FSL Harvard-Oxford Subcortical Atlas and *average subcortical* includes *bilateral accumbens*, *amygdala*, *caudate*, *hippocampus*, *pallidum*, *putamen*, and *thalamus*. The *gray matter* mask was defined as SPM's probabilistic gray matter map thresholded at $p < 0.5$.

As shown in **Fig. 4a-d**, SFS identifies advantages for dynamic fidelity in increasing temporal resolution, as well as the costs and benefits associated with increasing the number of head-coil channels vs. field strength in order to recover signal loss from accelerated acquisition. In general, the ultra-dense head-coil strategy employed by *Acquisition B* optimizes dynamic fidelity in cortical regions, whereas the ultra-high-field strategy employed by *Acquisition C* optimizes dynamic fidelity in subcortical regions. TSNR provides a very different story: showing the greatest stability in *Acquisition A*, diminished performance in *Acquisition B*, and the worst performance in *Acquisition C*. Which strategy is ideal, for any particular study, therefore depends critically upon the scientific questions to be asked: not only with respect to the regions of interest implicated, but also the types of analyses to be performed.

## Discussion

Functional neuroimaging has ushered in a new era of brain research, in which task-free fluctuations play an increasingly large role. As such, we need to reconsider whether fMRI optimization paradigms that rely solely on maximizing stability might actually be leading us astray, by failing to functionally dissociate fluctuations underlying signal versus those underlying noise. Here we propose a new measure—*Signal Fluctuation Sensitivity* (*SFS*)—that distinguishes between neurobiologically-relevant fluctuations of interest, and nuisance fluctuations due to physiological or scanner artifact. We demonstrate that SFS positively—and tSNR negatively—correlates with dynamic fidelity in a dynamic phantom, as well as with the detection power of local and long-range functional networks in humans, across three sets of representative acquisition parameters independently optimized for human fMRI studies.

Our design of the dynamic phantom was motivated by the need to rigorously test the fidelity of fMRI time-series to its known dynamic inputs, which would be otherwise impossible using either a static phantom or human data. While we could have simulated input-output fidelity in the presence of physiological and scanner noise, models can be susceptible to bias and often over-simplify the complexities of fMRI noise[27,28]. The empirical approach defined here captures actual scanner noise, and thus is more accurate in evaluating the utility of SFS to human neuroimaging. One of the most challenging aspects of our phantom's design from an engineering standpoint was the need to avoid motion artifacts for a machine with rotating elements. While the phantom's inner cylinder is programmed to move only between the scanner's intermittent radio frequency pulses, residual motion artifact would have a devastating impact on SFS values, because it would preferentially affect $\sigma_{ROI}$ from the inner (rotating) compartment, while bypassing $\sigma_{nuisance}$ from the outer (static) compartment. The fact that the slices from which we acquired data did not show characteristic motion artifact (clearly visible during slices that, for testing, were deliberately acquired during motion), that SFS values for the phantom correlated with dynamic fidelity (as shown in Online Methods, motion artifact corrupts fidelity), and that the relationships observed for dynamic fidelity were supported by human connectivity data, provides assurance with respect to the integrity of the phantom's design. Future validation of SFS would benefit from a biological ground truth for measurement of dynamic fidelity, using simultaneous inputs recorded from local field potentials and their associated hemodynamic responses, combined with outputs obtained from fMRI time-series.

For dynamic analyses, the structure of the SFS equation reflects the equally important need to optimize over signal amplitude (provided by the mean) and signal change (provided by the standard deviation). In our datasets, we found that the relationship between SFS versus tSNR and long-range connectivity was driven primarily by the standard deviation component of each equation (including only the standard deviation ratio component of the SFS equation, correlations were $r_{A,B,C} = 0.68$, $p = 4.34 \times 10^{-6}$ for N=36; unsmoothed), while the mean signal component (including only the mean amplitude of signal or the ratio of the mean amplitude to global amplitude) showed no statistical relationship to long-range connectivity strength ($r_{A,B,C} = -0.18$, $p = 0.31$; $r_{A,B,C} = 0.02$, $p = 0.90$). However, our fMRI data had minimal signal drop out in regions of interest, which is not always the case. While the mean amplitude of the signal did not play a role in evaluating our data set, nevertheless this term of the SFS equation should be retained in order to avoid assigning high SFS to areas of the brain that show signal loss.

In developing SFS for humans, one important decision is the optimal location for the acquisition of *nuisance* fluctuations. We chose *cerebrospinal fluid*, rather than surrounding air, white matter, or whole brain, because time-series from the cerebrospinal fluid contain the greatest proportion of nuisance variance of the three brain compartments[19], including motion, scanner variance, and some physiological effects. Moreover, unlike white matter[29] and the global signal, the eroded CSF masks used here are unlikely to contain neurobiologically-relevant fluctuations of interest.

In extending our phantom results to the brain, we faced the problem of what to look for as a measure of validation, since we lacked the phantom's advantage of known inputs. Thus, we used highly conservative and well-replicated connections in order to evaluate detection sensitivity for resting-state data: a measure of local-connectivity (*regional homogeneity – ReHo*[22]), a single voxel measure of resting-state signals (*amplitude of low frequency fluctuations - ALFF*[23]), and the *long-range connection* between two nodes within the default mode network (*medial prefrontal cortex* and *posterior cingulate cortex*)[25]. Both ALFF and ReHo have been widely used to study resting-state brain activity, with clinical applications to Parkinson's disease[30], Alzheimer's disease[31], and psychiatric illnesses[32]. Likewise, identification of the default mode network via long-range connectivity represents a fundamental finding in neuroscience[4,25], with direct implications for neurodevelopment and aging. Although the correlations for SFS and tSNR were both significant but of opposite sign, it is critical to note that they are not trivially inverses of one another. From a theoretical perspective, SFS and tSNR differ fundamentally in their dissociation of fluctuations of interest versus those of nuisance. From a practical perspective, **Fig. 4** demonstrates that the two measures provide qualitatively different mapping of optimization over the brain. Thus, we demonstrate that, by optimizing for *dynamic fidelity* rather than the current standard of *dynamic stability*, SFS can have direct practical applications for markedly increasing detection sensitivity of clinical neuroimaging results.

Although we have emphasized the application of SFS to correlational analyses due to their increasing prevalence within the field, it is important to note that other types of task-free analyses will also be much better served by optimization to SFS than tSNR. This category of analyses includes those based upon power spectra and complexity (e.g., *ALFF*, *power spectrum scale invariance*, *entropic analyses*, *spectral dynamic causal modeling*), which also are more highly sensitive to subtle dynamic features of the time-series than are traditional contrast-based analyses[*]. However, CNR and tSNR may still be useful and accurate measures in answering particular questions. Temporal SNR is a measure of signal stability that is proportional to field strength, voxel size, and sampling rate [13,33]; thus, in static phantoms, tSNR can be used to quantify and minimize scanner-related noise. If the primary aim of a study is to show contrast between two conditions then CNR, and not SFS, is correct. For task-free analyses, however, CNR is not directly measurable; thus, classical tSNR is normally cited as a surrogate[18,34]. It is in this case that using tSNR as guide will minimize, rather than maximize, detection sensitivity. As with so many zero-sum decisions in fMRI acquisition, it is important to realize that we optimize over one parameter at the expense of the other. Therefore, just as tuning of acquisition parameters benefits enormously from knowing *a priori* the region of interest to be targeted, knowing *a priori* the type of analysis to be performed will permit researchers decide whether to optimize for stability (tSNR) or for dynamics (SFS).

## Acknowledgements

This research was funded by the National Institutes of Health (NIDA-1R2DA03846701 LRMP) and the National Science Foundation (CBET-1264440 LRMP). The authors thank Jonathan Polimeni Ph.D., Sheeba Arnold Anteraper Ph.D., and Kin Foon Wong, Ph.D. for their assistance in scanning and troubleshooting, Randy Buckner Ph.D. for providing data from the Human Connectome Project for the analysis, as well as Jaime Shinsuke Ide, PhD, for data processing and discussions regarding the results.

---

[*] To illustrate the impact of signal fidelity on spectral methods, we calculated the power spectrum scale invariance (after detrending, with a full frequency range from 0 to the Nyquist limit of 0.5*1/TR) for time-series acquired with the dynamic phantom. As with connectivity analyses, SFS—but not tSNR—correlated with accuracy: the percentage difference (i.e., 'error'), between 'true' power-law slope β for the phantom and the 'measured' β values acquired from fMRI, decreased with higher SFS ($r = -0.67$; $p < 0.05$ for 9/9 scans) and increased with tSNR ($r = 0.66$; $p < 0.05$ for 8/9 scans).

Figure Captions

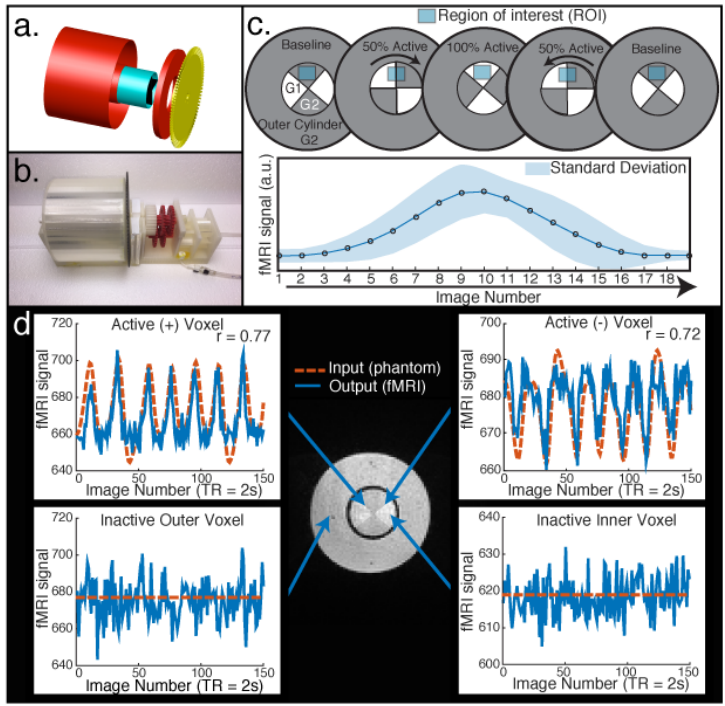

**Figure 1: The dynamic phantom produces tightly controlled changes in functional MRI signal, establishing a ground truth for quantifying dynamic fidelity of scanner outputs to signal inputs.** (*a, b*) The dynamic phantom uses concentric cylinders filled with agarose gels. The inner cylinder is coupled to an fMRI-compatible pneumatic motor and fiber optic feedback system. (*c*) The inner cylinder is longitudinally compartmentalized into four chambers. One of two calibrated agarose gels with different concentrations is contained in each; the gels are in direct contact. The outer cylinder contains a single agarose gel. Because magnetic susceptibility changes as a function of agarose concentration, precisely timed rotation of the inner cylinder between images creates a 'gradient' effect, in which different proportions of each agarose compartment pass through—and are averaged over—a region of interest. Motion across the 'gradient' thus is capable of producing smooth dynamic changes in fMRI signal (bottom panel of C). (*d*) The top two panels demonstrate "active" voxels within the inner cylinder of the phantom along the gel-gel interfaces; these voxels exhibit strong input-output fidelity. The bottom two panels show that the inactive outer cylinder and inactive inner cylinder voxels are indistinguishable. For validation of phantom performance, a simple event-related design is pictured in D. During the phantom scanning for SFS experiments, the phantom utilized a more complex input mimicking human resting-state fluctuations (**Fig. 2a**).

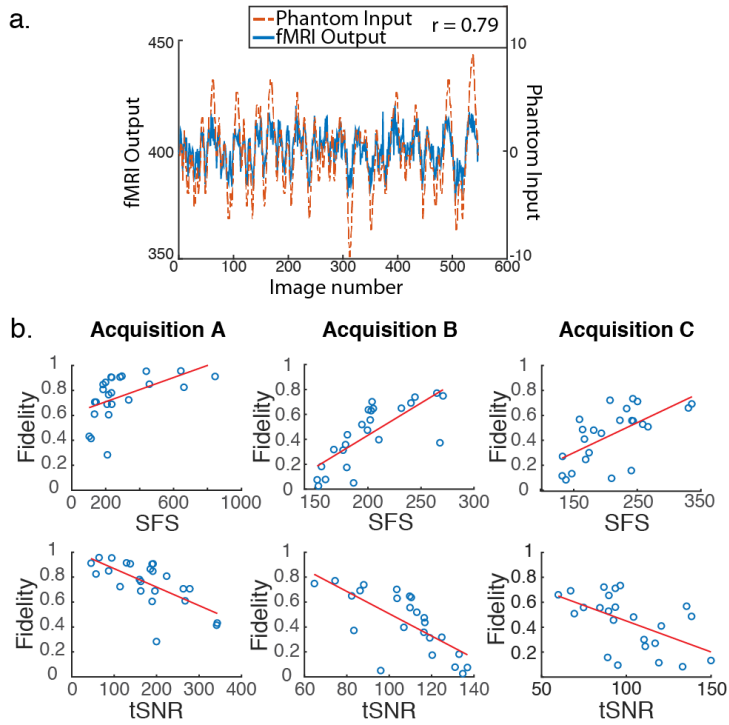

**Figure 2: Dynamic phantom results show dynamic fidelity positively correlates with signal fluctuation sensitivity (SFS) and negatively correlates with classical temporal signal to noise ratio (tSNR).** *(a)* To accurately mimic human resting-state fluctuations in the dynamic phantom, we utilized a complex pink-noise waveform as shown by the dotted line. The 10-minute input function originated from our previous neuroimaging data and was subsequently programmed into the phantom. The dynamic phantom inputs are derived from position tracking during rotation. A representative output fMRI signal is superimposed (*fMRI Output* axis), as acquired under *Acquisition B:* 3T magnet, 64 Channel head-coil, at TR = 1080ms (see **Table 1**). This waveform input was used for all nine phantom fMRI scans. *(b)* Input-output fidelity was positively correlated with SFS (median $r$ = 0.67, see **Table 2**) and negatively correlated with tSNR (median $r$ = -0.63, see **Table 2**). Groups presented here match the scanning parameters presented in the subsequent human data: acquisition A is a 3 Tesla magnet with a 32-channel headcoil (TR = 2000ms), acquisition B is a 3 Tesla magnet with a 64-channel headcoil (TR = 1080ms), and acquisition C is a 7 Tesla magnet with a 32-channel head coil (TR = 802ms). **Table 1** provides detailed acquisition parameters for each scan, while **Table 2** provides detailed results from all nine dynamic phantom scans.

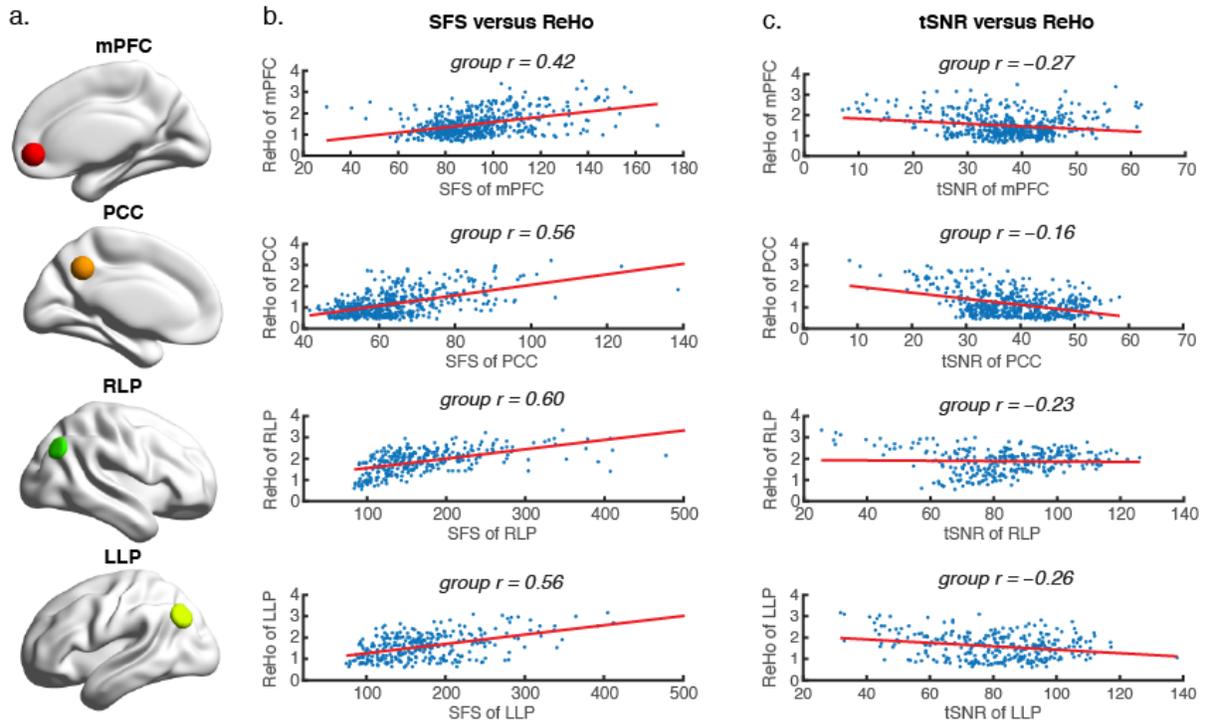

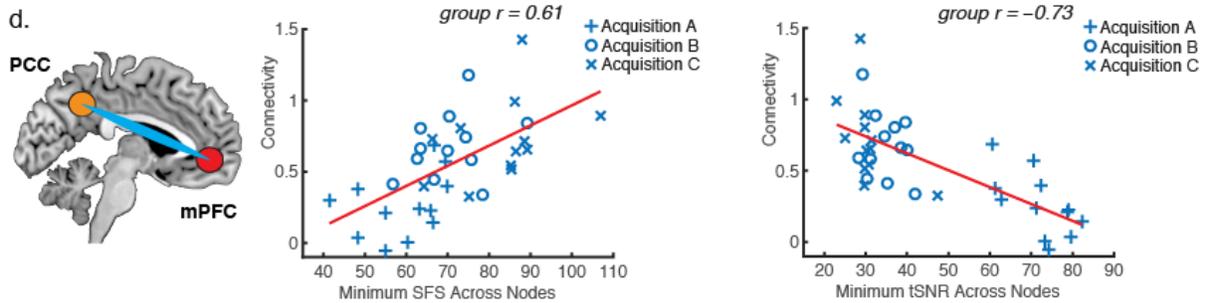

**Figure 3: Local and long-range functional connectivity across the default mode network positively correlates with SFS and negatively correlates with tSNR.** *(a)* We calculated SFS regional homogeneity (ReHo, a commonly used measure of neural synchrony in fMRI) for each individual subject across the *medial prefrontal cortex* (*mPFC*), *posterior cingulate cortex* (*PCC*), and *right and left lateral parietal lobes* (*RLP* and *LLP*). *(b-c)* Detection sensitivity for ReHo positively correlates with SFS and negatively correlates with tSNR (scatter plots shown for a single representative subject; group *r* for N=36). *(d)* We see that the same pattern occurs for long-range connectivity between default mode network regions medial prefrontal cortex (mPFC) and posterior cingulate cortex (PCC). As spatial smoothing artificially increases ReHo by producing correlations between contiguous voxels, shown data are unsmoothed.

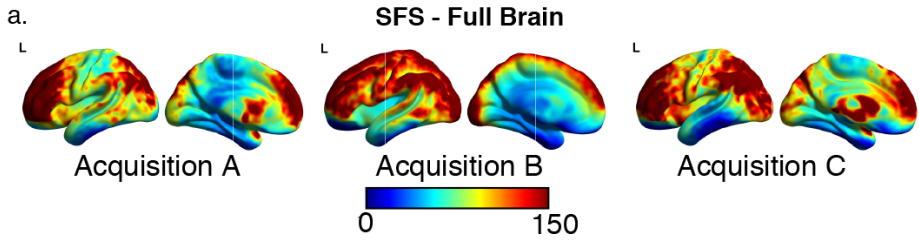
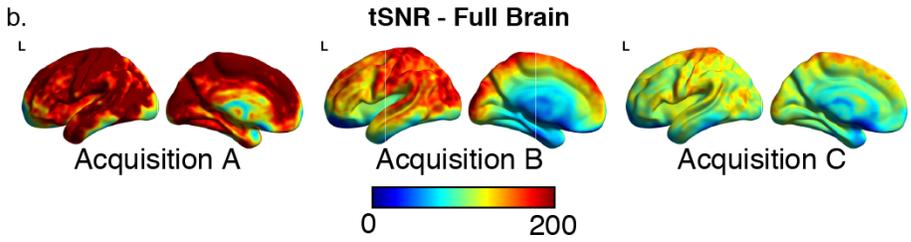
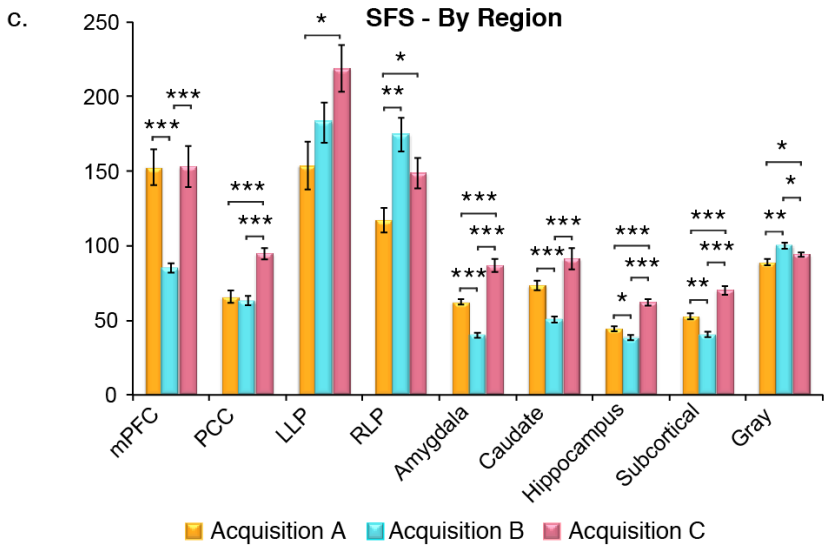
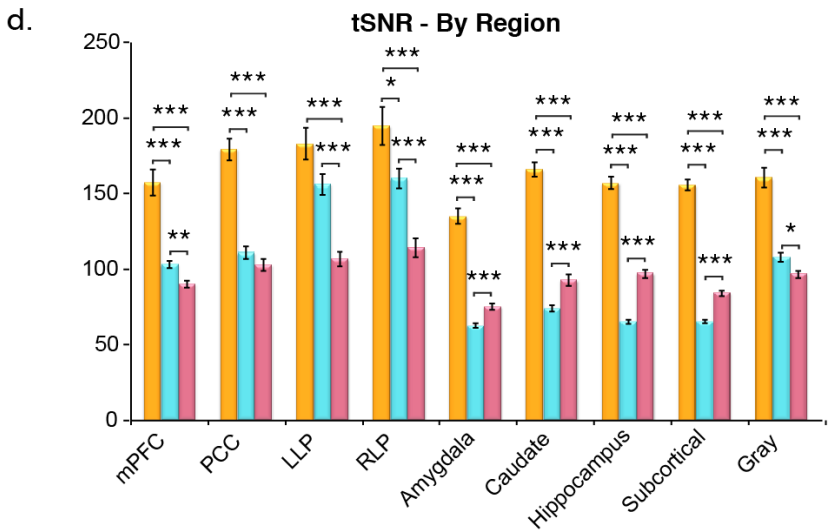

**Figure 4: SFS distributions across the brain illustrate qualitative differences in sensitivity between acquisition strategies.** As before, acquisition A is a 3 Tesla magnet with a 32-channel head coil (TR = 2000ms), acquisition B is a 3 Tesla magnet with a 64-channel head coil (TR = 1080ms), and acquisition C is a 7 Tesla magnet with a 32-channel head coil (TR = 802ms). *(a)* Full brain SFS maps for each acquisition demonstrate that cortical (especially prefrontal and parietal/visual) SFSs are robust across all acquisitions. Acquisition B shows more uniform cortical SFS than A or C, while acquisition C shows greater subcortical SFS than A or C. *(b)* SFS values across acquisition strategies averaged within several regions, including the default mode network, subcortical regions, and gray matter. In general, SFS was maximized in cortical regions for acquisition B and subcortical regions for acquisition C. *(c)* Acquisition A demonstrated the highest tSNR for all regions, followed by acquisition C and acquisition B. Values were derived from preprocessed and smoothed resting-state data (n = 12 per group, 5 minutes of data).
*$p$ < 0.05, **$p$ < 0.01, ***$p$ < 0.001 (Wilcoxson rank sum test).

Tables

**Table 1: Acquisition parameters for the nine Dynamic phantom scans.** We tested three scanners at three TRs with the dynamic phantom. Highlighted rows indicate scans for which we collected corresponding human data, in which Scan 1 corresponds to *Acquisition A*, Scan 5 corresponds to *Acquisition B*, and Scan 9 corresponds to *Acquisition C*. All sequences utilized interleaved acquisition.

| Scan | Main Field | Head Coil | TR (ms) | TE (ms) | SMS | iPAT | Flip Angle | Bandwidth (Hz/Px) | Resolution (mm) | Slice Gap (mm) | Slices |
|---|---|---|---|---|---|---|---|---|---|---|---|
| 1 | 3T | 32 Ch | 2000 | 30 | 3 | 2 | 75° | 1860 | 2x2x2 | 0.2 | 69 |
| 2 | 3T | 32 Ch | 1080 | 30 | 4 | 2 | 60° | 1860 | 2x2x2 | 0.2 | 60 |
| 3 | 3T | 32 Ch | 802 | 30 | 5 | 2 | 33° | 1860 | 2x2x2 | 0.2 | 55 |
| 4 | 3T | 64 Ch | 2000 | 30 | 2 | 2 | 85° | 2840 | 2x2x2 | 0 | 62 |
| 5 | 3T | 64 Ch | 1080 | 30 | 4 | 2 | 60° | 2840 | 2x2x2 | 0 | 68 |
| 6 | 3T | 64 Ch | 824 | 30 | 5 | 2 | 55° | 2840 | 2x2x2 | 0 | 65 |
| 7 | 7T | 32 Ch | 2010 | 20 | 2 | 2 | 33° | 2264 | 2x2x1.5 | 0 | 86 |
| 8 | 7T | 32 Ch | 1010 | 20 | 4 | 2 | 55° | 2264 | 2x2x1.5 | 0 | 84 |
| 9 | 7T | 32 Ch | 802 | 20 | 5 | 2 | 33° | 2368 | 2x2x1.5 | 0 | 85 |

**Table 2: Dynamic fidelity in all nine dynamic phantom scans was positively correlated with SFS and negatively correlated with tSNR.**
*n.s.

| | TR (ms) | 3T, 32 Ch | 3T, 64 Ch | 7T, 32 Ch |
|---|---|---|---|---|
| | | **Correlation with Dynamic Fidelity** | | |
| **SFS** | ~2000 | 0.51 ($p=0.010$) | 0.69 ($p=2.1\times10^{-4}$) | 0.54 ($p=6.6\times10^{-3}$) |
| | ~1080 | 0.71 ($p=1.1\times10^{-4}$) | 0.76 ($p=1.8\times10^{-5}$) | 0.69 ($p=1.9\times10^{-4}$) |
| | ~802 | 0.67 ($p=3.7\times10^{-4}$) | 0.49 ($p=0.014$) | 0.63 ($p=9.0\times10^{-4}$) |
| **tSNR** | ~2000 | −0.67 ($p=3.2\times10^{-4}$) | −0.64 ($p=7.0\times10^{-4}$) | −0.44 ($p=0.030$) |
| | ~1080 | −0.87 ($p=4.6\times10^{-8}$) | −0.72 ($p=8.3\times10^{-5}$) | −0.58 ($p=3.0\times10^{-3}$) |
| | ~802 | −0.63 ($p=1.1\times10^{-3}$) | −0.27* ($p=0.20$) | −0.54 ($p=6.2\times10^{-3}$) |

Online Methods

**Design of Dynamic Phantom**

*Summary of strategy:* We designed a dynamic phantom that is fully automated, capable of producing complex waveforms detected by fMRI, and contains no paramagnetic materials. The basis for the signal is the fact that the magnetic susceptibility of agarose gels is concentration dependent[20], in which higher concentrations produce lower fMRI signal. By varying the concentration of agarose gel present within a voxel over time, the dynamic phantom produces changes in T2* that can be tuned to amplitudes typically seen with BOLD in humans; the phantom can be programmed to simulate both task-based (stimulus-driven, as shown in **Fig. 1d**) and resting-state (pseudo-random fluctuations, as shown in **Fig. 2a**) BOLD-like signals. With known inputs, the relationship between the signal produced (by the dynamic phantom) and the signal detected (by the fMRI scanner) can be rigorously quantified as a measure of *dynamic fidelity*.

The dynamic phantom is composed of calibrated agarose gels housed within two concentric cylinders. The outer cylinder contains a *baseline* gel, while the inner cylinder is longitudinally divided with both (*i*) a baseline gel matching the outer cylinder and (*ii*) an *active* gel with slightly lower concentration of agarose (**Fig. 1**). The longitudinally divided inner cylinder produces dynamic fMRI signal via rotation about its long axis. We developed a novel fMRI-compatible pneumatic motor to drive rotation of the inner cylinder, while the outer cylinder remains motionless. The position of the inner cylinder is continuously monitored with a fiber optic feedback system, and the device is operated from the fMRI control room with a microcontroller. Compressed air drives rotation of the inner cylinder, and monitoring of the phantom occurs through plastic fiber optic cables, which run between the scanner and control room. Thus, the dynamic phantom is comprised of two main systems: 1) the *scanned phantom*, consisting of two concentric cylinders and supports, a plastic gearbox, tubing, and fiber optic cables, and 2) the *control unit*, consisting of a microcontroller, compressor, and circuit board. The description of the design will be broken down within these two systems and their interface. A MATLAB toolbox with all computational and visualizations already implemented in a user-friendly manner can be found at: http://www.lcneuro.org/software-and-instrumentation/.

System 1: Scanned Phantom

*Phantom housing:* We used AutoCAD (AutoDesk, Inc) to design a cylinder-within-cylinder phantom (Fig. 1). The inner cylinder contains four compartments, divided longitudinally. All custom phantom parts were printed with a Makerbot 3D printer with non-pigmented polylactic acid filament (Makerbot, Inc, Brooklyn, NY). The volume of the outer cylinder was 600 mL, while the volume of the inner cylinder was 150mL.

*Agarose gels and materials justification:* We chose agarose as a contrast medium because of its relative ease of use, flexibility in preparation, and physical stability. The use of agarose in phantom construction has been validated throughout the literature, and it is shown to be homogenous with respect to MR relaxation properties[20,35]. The outer cylinder was filled uniformly with 2.27% agarose. The inner cylinder was filled with 2.21% and 2.27% agarose gels (Figure 1B). No dividing materials were used, i.e., the gels were in direct contact. We used a "baseline" agarose gel concentration approximately matching previous agarose phantom development[20]. We then calibrated the 'active' agarose gel concentration by empirically measuring fMRI signal intensity at 3T for agarose concentrations between 1.5% and 3.0%, and chose the concentration that produced an approximate 1% signal change in a 3T 12 channel MRI during a simulated event-related design. Gels were degassed with a vacuum chamber.

Interface between System I and System II

*Control and automation:* To achieve automated rotation of the inner cylinder, we designed and fabricated a fully fMRI compatible pneumatic motor system. The motor consists of a compressor, valves, manifold, tubes, dual fans, and a gearbox. An air compressor is placed in the control room of the fMRI center; input pressure is set to 40 pounds per square inch at 1.9 cubic feet per minute. Plastic tubing guides the compressed air through a splitter and into two Arduino controlled solenoid valves (SparkFun Electronics, Boulder CO). Compressed air leaves the two independent valves and is guided through two tubes into the scanner bed. The compressed air is released from the pairs of tubes via pneumatic connectors, resulting in high velocity airflow. Depending on which valve is open, this airflow powers one of two fans; these fans are coupled to a gearbox and spin in opposing directions. The dual fan setup allows the gearbox to be driven in either direction and also allows precise braking. The rapid rotation of the fans is stepped down and torque is increased via five 3:1 compound gears, resulting in a step down ratio of 243:1. The gearbox ultimately interfaces with the inner cylinder and optical interruption disk to produce pneumatically controlled rotation. The outer cylinder does not rotate.

*Fiber optic feedback:* We designed a fiber-optic feedback system using plastic fiber-optic cables, an LED light source, a photodiode, and an interrupter disc. An Arduino microcontroller powers a high-powered 10 mm LED (SparkFun Electronics, Boulder CO), which is coupled with a 1.5 mm diameter fiber optic cable (Thor Labs, NJ). The first cable guides light from the LED source within control room to the scanner bed through a waveguide. The fiber optic cables are positioned opposite each other and spaced 5mm apart, such that as the inner cylinder rotates, the interrupter disk (3mm thickness) will intermittently block light transmission between the two cables. The interrupter disc has 60 teeth, corresponding to ~6° of rotation per interrupt. The second fiber optic cable receives light and is fed back to a photo-diode on the microcontroller. As the interrupter disc spins, the photo-diode receives differential intensity readings. The microcontroller then displays the interruption count as a live feed at each TR. We calibrated the phantom to traverse an average of one interrupt per image. Prior to each fMRI scan, the device performs a self-calibrating procedure to ensure optimal position encoding regardless of ambient light.

System II: Control Unit

*Arduino microcontroller and fMRI communication:* TR signals are sent to the Arduino via USB input from the fMRI. To properly calibrate the phantom rotation and avoid motion artifacts in regions of interest, we ran a simple EPI acquisition (TR = 2, TE = 30ms, 25 slice) in which the phantom began rotation just after the start of each TR, and examined each slice for motion artifacts. We found that motion artifacts occurred when the phantom rotated during or before a slice was acquired, whereas slices acquired before the phantom rotates within a TR contained no noticeable artifact. Because the phantom rotates in plane with the image, no material leaves or enters the imaging slice; this feature avoids potential spin history artifacts. Therefore, if the phantom is programmed to begin rotation towards the end of a TR (after a sufficient number of slices have been acquired) and to stop rotation *just before* the next TR, motion artifacts are mitigated (see Fig. 1d for a representative time-series acquired with the dynamic phantom). Empirical testing with this design indicated that the phantom should begin rotation 650ms prior to each TR, and stop ~100ms before the TR. Thus, for TR = 2s, the dynamic phantom begins rotation at 1500ms and ends at 1900ms; for TR = 1080, we began rotation at ~600ms and ends at ~980ms; for TR = 802ms, rotation begins at ~300ms and ends at ~700ms. This strategy produced minimal motion artifacts in images of the center of the phantom, where inactive inner cylinder voxels (which experience motion) and inactive outer cylinder voxels (which experience no motion) showed no significant differences in standard deviation ($p = 0.89$, rank sum test).

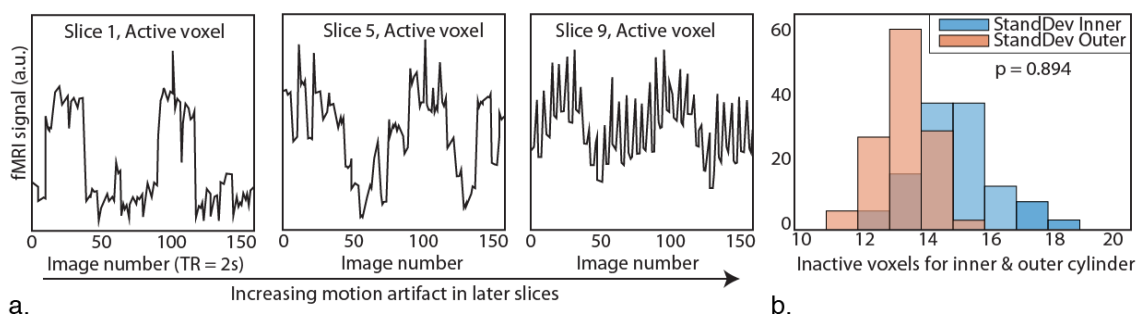

**Online Methods Figure 1: Motion artifacts during rotation vs. slice.** The dynamic phantom rotates between 3-6° between TRs. Rotation is coupled with TR acquisition through a microcontroller, and is tightly controlled with a brake. For illustrative purposes, we show here that slices acquired before rotation *(a)* are subject to considerably less spiking than slices acquired during rotation *(b)* and after rotation is completed *(c)*. As shown in **Fig. 1d**, we optimized our rotation/braking scheme such that **inactive** inner and outer cylinder voxels contain no significant differences in standard deviation for slices of interest (rank sum test)*(d)*.

*Arduino software:* The dynamic phantom is controlled with an Arduino Mega (www.Arduino.cc). We developed all software in-house. The phantom can operate in three distinct modes: 1) *stimulus-driven* (for simulation of task-designs), 2) *guided-mode* (for simulation of resting-state), and 3) *static*. For this experiment, the phantom utilized *guided mode*, for which the user preprograms the interruption destination for each image. This allowed for the production of specific time-series, such as a pink-noise time-series equivalent to those produced by resting-state fMRI (Fig. 2a).

**Using the dynamic phantom to test SFS and tSNR**

*Acquisition parameters:* We scanned the dynamic phantom in three separate fMRI scanners. Detailed scan parameters are listed in **Table 1**. The three scanners utilized in this phantom study represent the following: (*i*) a 3T Siemens MRI with 32-channel head-coil (McGovern Institute for Brain Research, Massachusetts Institute of Technology), (*ii*) a 3T Siemens MRI with 64-channel head-coil (Human Connectome Scanner—Martinos Center for Biomedical Engineering, Massachusetts General Hospital), and (*iii*) a 7T Siemens MRI with 32-channel head-coil (Martinos Center for Biomedical Engineering, Massachusetts General Hospital). For each scanner, we tested three sampling rates, representing typical time-resolution for fMRI studies (TR=2000-2010ms), increased time-resolution acquired for the Human Connectome Project (TR=1010-1080ms), and ultra-fast imaging paradigms (TR=802-824ms). Thus, we performed a factorial study (three scanners and three sampling rates each) with the Dynamic phantom, for a total of nine scans (**Table 1**), each 10 minutes long. For both 3T scanners, we performed standard shimming; due to dramatically increased susceptibility artifacts at 7T, we utilized a partial shim centered on the inner cylinder of the phantom. Visual inspection of the resulting images, as well as correlations between the Dynamic phantom inputs and fMRI outputs, confirmed data quality.

*Statistical analyses:* While most human fMRI data undergoes significant preprocessing, for the dynamic phantom we used raw data after implementing only voxel-wise trend removal (linear and quadratic) to remove scanner drift, and no further temporal preprocessing, in order to characterize dynamic fidelity as transparently as possible. For the *region of interest* (*ROI*) fluctuations, we extracted the average time-series from the four quadrants of the inner cylinder (corresponding to the four chambers) with an automated masking procedure using MATLAB software developed in-house. We repeated this for six slices positioned in the center of the phantom (n = 24 time-series per scan). For the *nuisance* fluctuations, we extracted the time-series from the outer cylinder of the phantom, which does not activate. We then computed quadrant-wise SFS based on the definition:

**Online Methods Eqn. 1**

$$SFS_{dynamic\ phantom\ quadrant} = \frac{<\mu_{inner\ cylinder}>_{quadrant}}{<\mu_{global}>} \times \frac{<\sigma_{inner\ cylinder}>_{quadrant}}{<\sigma_{outer\ cylinder}>}$$

In the first term, the numerator consists of the mean signal ($\mu$) of an averaged time-series over each of the four dynamic phantom quadrants (*quadrant*). For the denominator, we average over signal for the entire phantom (*global*). The first term ensures that SFS decreases for regions with signal drop out, while remaining unit-less (as with tSNR). In the second term, the numerator consists of the mean standard deviation ($\sigma$) of an averaged time-series over each of the four dynamic phantom quadrants. For the denominator, we average over $\sigma$ from a region in which signals are not expected, but in which physiological, scanner, and motion artifacts are still present. In this case, we use the outer cylinder, which is static. In order to avoid biasing values for standard deviation due to differences in the number of voxels between inner quadrants and outer cylinder, we averaged time-series in the outer compartment over the same number of voxels used to average time-series in each of the inner quadrants. We computed standard deviations for each of these inner quadrant-sized (39 voxel) averaged time-series, and then averaged across those standard deviations to produce the standard deviation for the entire outer cylinder (i.e., the denominator of the second factor). In order to more easily compare SFS with tSNR, we scale them comparably by multiplying SFS values by 100. TSNR was computed as the mean for the averaged time-series over each of the four dynamic phantom quadrants, divided by its standard deviation (after detrending). Dynamic fidelity was computed as the correlation between inputs (dynamic phantom user-defined function) and outputs (fMRI time-series). We then computed the correlation between fidelity and both SFS and tSNR for each of the 24 time-series per scan.

**Human scanning**

*Acquisition:* In an effort to represent a wide variety of task-free scanning paradigms, we analyzed three sets of human data (n = 12 subjects each) collected with the same acquisition parameters utilized for the phantom studies, but using only the time-resolutions previously optimized for each study (Table 1). Thus, *Acquisition A* refers to the 3T fMRI with 32-channel head-coil and a TR= 2000ms; *Acquisition B* refers to the 3T fMRI with a 64-channel head-coil and a TR=1080ms, and *Acquisition C* represents the 7T fMRI with a 32-channel head-coil with a TR=802ms. *Acquisition A* lasted 5 minutes, while *Acquisition B* (originally 6.2 minutes) and *Acquisition C* (originally 10 minutes) data were truncated to match this duration. Anterior to posterior phase encoding and interleaved acquisition were used in all scans. For *Acquisition A*, we acquired whole-brain T1-weighted structural volumes using a conventional MPRAGE sequence with the following parameters: TR = 2530ms, TE = 3.39ms, TI = 1100ms, flip angle = 7°, voxel size = 1 x 1 x 1.3mm. Conventional B0 field maps derived from phase differences between gradient echo images acquired at TR = 4.22 and 6.68ms were also acquired (TR = 584ms, flip angle = 55°, voxel size = 2 x 2 x 2 mm, slice gap = 0.2mm, 69 slices). For *Acquisition B*, we also acquired whole-brain T1-weighted structural volumes using a conventional MPRAGE sequence with TR = 2530ms, TE = 1.15ms, TI = 1100ms, flip angle = 7°, 1mm isotropic voxel size. For *Acquisiton C*, we acquired whole-brain T1-weighted structural volumes using a multi-echo MPRAGE (MEMPRAGE) sequence with four echoes and the following protocol parameters: TR = 2530ms, TE1 = 1.61ms, TE2 = 3.47ms, TE3 = 5.33ms, TE4 = 7.19ms, TI =1100ms, flip angle = 7°, 1mm isotropic voxel size. Conventional B0 field maps derived from phase differences between gradient echo images acquired at TE = 4.60 and 5.62ms were also acquired (TR = 723ms, flip angle = 36°, voxel size = 1.7 x 1.7 x 1.5 mm, 89 slices).

All subjects were age matched ($\mu_A$ = 25.6 ± 3.7; $\mu_B$ = 23.3 ± 4.2; $\mu_C$ = 25.6 ± 3.4, *p* = 0.35, Kruskall-Wallis test). There were no significant differences in motion across the three groups (maximum absolute translation p = 0.60, maximum absolute rotation p = 0.96, mean root mean square (RMS) motion p = 0.10, maximum RMS motion p = 0.27, Kruskall-Wallis test). All participants were instructed to lie quietly with eyes open in the scanner, orienting to a fixation cross, without moving for the duration of the scan. We removed the first ten seconds of data for all datasets.

*Preprocessing:* We followed the standard SPM 8 pipeline for realignment, co-registration to a structural image, and normalization to Montreal Neurological Institute (MNI) space. Co-registered structural images were segmented into probabilistic maps of gray matter, white matter, and CSF using SPM's New Segment tool. Where noted, we utilized a 4-mm (2 voxel) FWHM Gaussian smoothing kernel. As per standard practice for fMRI analyses, we performed slice time correction only on *Acquisition A* data, since the 2000ms sampling rate was considerably slower than those of the other two scanners. We performed field map correction on *Acquisitions A and C* (distortion correction scheme was performed on *Acquisition B* immediately following image acquisition). Scrubbing was performed to remove the influence of motion, with scan-to-scan global signal deviation from the mean > 3 and scan-to-scan composite motion > 0.5 mm as thresholds for removal[36]. The mean percentage of data points removed between all three groups was 1.97%, with no subjects having more than 9% of data scrubbed. To assess the impact of spatial smoothing, we computed all of our measures on both unsmoothed and smoothed data, both of which underwent each of the other preprocessing steps listed here.

**Computation of SFS, tSNR, ALFF, ReHo and Long-Range (mPFC-PCC) Connectivity**

We used MATLAB to compute voxel-wise SFS according to:

**Online Methods Eqn. 2**

$$SFS_{voxel} = \frac{\mu_{default\ mode\ network\ ROI}}{<\mu_{global}>} \times \frac{\sigma_{default\ mode\ network\ ROI}}{<\sigma_{cerebrospinal\ fluid}>}$$

In the first term, the numerator consists of the mean signal ($\mu$) of a time-series acquired from a voxel in the *region of interest* (*ROI*) in the default mode network, as defined below. For the denominator, we average over all voxel-specific signal for the entire brain (*global*). The first term ensures that SFS decreases for regions with signal drop out, while remaining unit-less (as with tSNR). In the second term, the numerator consists of the standard deviation ($\sigma$) of a time-series acquired from the voxel of interest in the default mode network, as defined below. For the denominator, we average over all voxel-specific $\sigma$ from a region in which BOLD signals are not expected, but in which physiological, scanner, and motion artifacts are still present (*nuisance*). Prior work suggests that time-series obtained from cerebrospinal fluid (CSF) meet criteria for the *nuisance* denominator[19]. SFS for a region of interest is then computed by averaging voxel-specific SFS values over all voxels in the region ($SFS_{ROI} = <SFS_{voxel}>_{ROI}$). We additionally computed voxel-wise tSNR as the mean for each voxel's time-series divided by its standard deviation. In order to more easily compare SFS with tSNR, we scale them comparably by multiplying SFS values by 100.

    For SFS, standard deviations of the cerebrospinal fluid voxels (*nuisance fluctuations*) were computed using an eroded probabilistic map of CSF (SPM8 segmented map of CSF thresholded at 70%), to ensure minimal contributions from neural sources. To avoid distorting time-series dynamics by averaging them, standard deviations were computed for each voxel in the nuisance ROI, with voxel-based values averaged for the ROI.

    Mean global signal included the entire brain (conjunction of gray matter, white matter, and cerebrospinal fluid, thresholded at 70%). Mean values and standard deviations for each voxel were acquired before confound correction, but after SPM8 preprocessing and scrubbing.

    Prior to functional connectivity analyses, we performed further regression of nuisance variables (confound-correction). This included detrending, regression of mean CSF and white matter signals (white matter map thresholded at 70%), and regression of six motion parameters from the realignment step. Finally, we performed temporal band-pass filtering in the 0.01-0.1 Hz range using 5[th] order Butterworth filter.

    Both *amplitude of low frequency fluctuations* (*ALFF*) and *local synchronization of neighboring voxels* (*regional homogeneity* or ReHo: 27-voxel KCC-ReHo) were computed from confound-corrected data, using the REST toolbox[37]. Resulting subject-specific voxel-wise ReHo and ALFF maps were standardized by dividing each voxel's value by the mean value of the whole brain.

    To test whether SFS or tSNR were predictive of these established resting-state measures, we computed within-subject correlations between (*i*) SFS and ALFF, (*ii*) SFS and ReHo, (*iii*)

tSNR and ALFF, and (*iv*) tSNR and ReHo for voxels belonging to the well-established *default mode network* (DMN) regions: *medial prefrontal cortex* (*mPFC*), *posterior cingulate cortex* (*PCC*), and *left and right lateral parietal cortices* (*LLP* and *RLP*). These regions were defined as 10-mm radius spheres centered on previously established coordinates[26], intersected with an SPM8 brain mask to ensure only brain voxels were included[26]. For the extraction of ROI-based SFS and tSNR values, we used the four aforementioned DMN masks, as well as a probabilistic gray matter mask from SPM8 ($P > 50\%$). We obtained subcortical ROI masks from bilateral regions included in FSL Harvard-Oxford subcortical atlas (thresholded at 50%).

As a measure of long-range mPFC-PCC connectivity strength we used Fisher-z transformed correlations coefficients between mean time series extracted for mPFC and PCC. To test whether SFS or tSNR were predictive of the mPFC-PCC connection strength, we computed between-subject correlations (N = 36) between the minimum SFS or tSNR for each mPFC-PCC pair and connectivity strength. The decision to use the SFS or tSNR value for the system as a whole based upon the minimum value is intuitively based upon the intuition that for networks that include two or more nodes, signal for the network as a whole can only be as strong as that of its weakest node. However, this intuition is not completely accurate; it is only a better solution than the next easiest option, which is to take the mean.

**Estimating the Impact of Noise on Correlations**

While some types of analyses (ReHo) are calculated from a single node, for other types of analyses (e.g., long-range connectivity, dynamic causal modeling) it may be desirable to optimize over the system/circuit as a whole. In order to do so, it is necessary to calculate the 'mutual' tSNR or SFS of a multiple-node system, for which each node may have different values. While one option would be to calculate the tSNR or SFS for each node, and then average between them, it turns out that this approach underestimates the impact of noise on the statistics.

To address the general question of how noise affects the Pearson correlation coefficient $r(x,y)$, we start with its definition:

**Online Methods Eqn. 3**

$$r(x,y) = \frac{1}{n-1} \sum_{i=1}^{n} \left( \frac{x_i - \bar{x}}{s_x} \right) \left( \frac{y_i - \bar{y}}{s_y} \right).$$

Thus:

**Online Methods Eqn. 4**

$$\bar{x} = \frac{1}{n} \sum_{i=1}^{n} x_i \text{ and } s_x = \sqrt{\frac{1}{n-1} \sum_{i=1}^{n} (x_i - \bar{x})^2} ,$$

and analogously for $\bar{y}$ and $s_y$.

Let us assume that both datasets *x* and *y* consist of correlated data and uncorrelated noise. We can therefore write:

**Online Methods Eqn. 5**

$$x_i = x_{Ci} + x_{Ni}$$

If we define $SNR_x$ as $x_{Ci}/x_{Ni}$ then

**Online Methods Eqn. 6**

$$x_i = x_{Ci}\left(1 + \frac{1}{SNR_x}\right)$$

At the same time $s_x$ calculates in the following way, assuming that $x_C$ and $x_N$ are uncorrelated:

**Online Methods Eqn. 7**

$$s_x^2 = s_{Cx}^2 + s_{Nx}^2 = s_{Cx}^2 + \frac{s_{Cx}^2}{SNR_x^2} = s_{Cx}^2\left(1 + \frac{1}{SNR_x^2}\right)$$

Therefore, if $r(x_C,y_C)=r_C$, and we substitute $x_i - \bar{x} = (x_{Ci} + x_{Ni} - \overline{x_{Ci}} - \overline{x_{Ni}})$ (and equivalently for the *y* part) in the definition of *r* above, then:

**Online Methods Eqn. 8**

$$r(x,y) = \frac{r_C s_{Cx} s_{Cy}}{s_x s_y} = \frac{r_C}{\sqrt{1 + \frac{1}{SNR_x^2}}\sqrt{1 + \frac{1}{SNR_y^2}}}$$

We can illustrate the practical impact of noise on measured correlations between time series by assuming the existence of two signals with perfect correlation ($r_C$ =1), each of which is subjected to different levels of noise (provided by tSNR values that match the variation within the literature: 4.42-280[17]). As shown by **Online Methods Eqns. 3-8**, if Node 1 has SNR of 4.42, and Node 2 has SNR of 280, the *r*-value will actually decrease from 1 to 0.975. On the other hand, averaging the two tSNR values provides an adjusted *r*-value of ~1 (0.999951). Obviously, our approach in taking the lowest of the tSNR values for all nodes is also inaccurate, overestimating the impact of the noise to *r*=0.951. However, for the purpose of optimization, it makes more sense to err on the side of being conservative, and thus we take the lower value (overestimating the impact of noise) rather than the average (underestimating the impact of noise).

Unsurprisingly, even without considering the compensatory fitting and averaging steps typically employed in contrast-based analyses, correlations are more sensitive to distortions of the frequency spectrum than are traditional contrast-based analyses. The purpose of dynamic fidelity, and therefore also of SFS, is to preserve the frequency spectrum. In order to do so, we require that time-series be linearly amplified. Let us assume that the measured BOLD signal has undergone amplification and also assume that this amplification is linear over some range of $t_2^*$s but that it is non-linear over the edges of the linear range (a sigmoidal shape for example). Since the $t_2^*$s vary over the brain regions of interest some voxels will be amplified linearly and some will not. Even if there were a perfect correlation between two brain regions, such a distortion would reduce the correlation, whereas for task-based designs this distortion would not be as much of a problem since only the difference between contrasts is important. To illustrate this, we can use a pink-noise power law time-series modeled upon our resting-state data (**Online Methods Fig.1**). We then transform the data using a sigmoidal curve to simulate a non-ideal amplifier as shown in **Online Methods Fig. 2**. The resulting transformation (**Online Methods Fig. 3**) lowers the correlation with the original time-series, from *r*=1 to *r*=0.8.

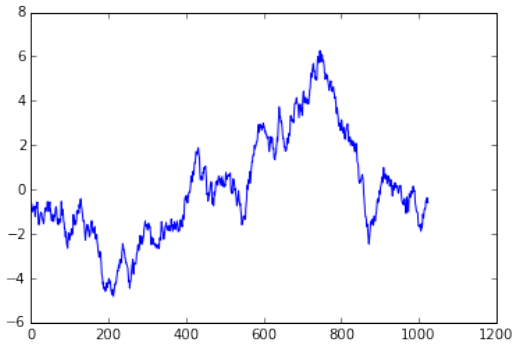

**Online Methods Fig. 1**: Simulated data set whose power spectrum obeys a power law with mean zero.

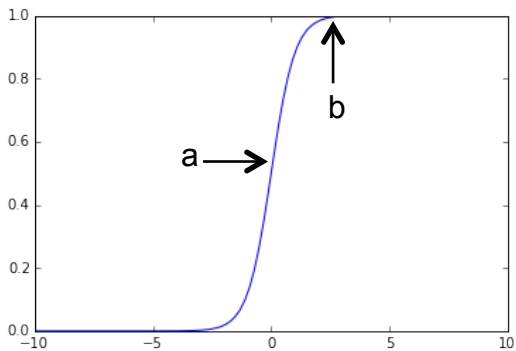

**Online Methods Fig. 2: Characteristic function of the non-linear amplifier.** To generate this curve we used a scaled version of the logistic function $1/(1+\exp(-2x))$. If the mean of the data is around 0, this amplifier is perfectly linear (*a*), but when the mean shifts beyond the linear range (*b*) this amplifier will distort the data as shown in **Online Methods Fig. 3.**

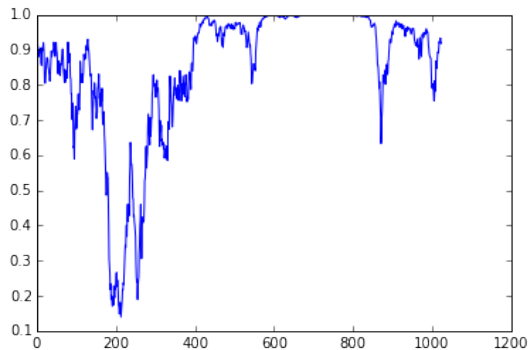

**Online Methods Fig. 3:** Data set in **Online Methods Fig. 1** has undergone non-linear amplification by shifting the mean up by 3 and transforming the data using the characteristic function in **Online Methods Fig. 2b**. These two data sets (original and transformed) maintain the power law but their Pearson correlation is reduced from $r$=1 to $r$=0.8.

The typical rule of thumb for optimizing fMRI is to set TE such that the signal amplifies at the center of the linear range. This is consistent with our aims, since **Online Methods Figs. 1-3** demonstrate that scanning in the nonlinear ends of the range will distort the time-series dynamics. Optimizing for SFS puts one at the center of the linear range (**Online Methods Fig. 2a**), where responses are maximized. However, optimizing over tSNR will always place one in the upper nonlinear location (**Online Methods Fig. 2b**), since it is the point at which the amplitude is highest and fluctuations are minimized.